\documentclass[12pt]{article}

\catcode`\@=11

\global\arraycolsep=2pt
\usepackage{amsbsy,amssymb,latexsym,amsfonts,amsmath}
\usepackage{graphicx,color}

\allowdisplaybreaks

\begin{document}

\vspace*{0.7cm}

\begin{center}
{ \Large Bootstrapping microcanonical ensemble in classical system}
\vspace*{1.5cm}\\
{Yu Nakayama}
\end{center}
\vspace*{1.0cm}
\begin{center}

Department of Physics, Rikkyo University, Toshima, Tokyo 171-8501, Japan

\vspace{3.8cm}
\end{center}

\begin{abstract}
We point out that the bootstrap program in quantum mechanics proposed by Han et al reduces to a bootstrap study of a microcanonical ensemble of the same Hamiltonian in the $\hbar \to 0$ limit. In the limit, the quantum mechanical archipelago becomes an extremely thin line for an anharmonic oscillator. For  a double-well potential, a peninsula in $E\le 0$ appears.
 
\end{abstract}

\thispagestyle{empty} 

\setcounter{page}{0}

\newpage

A problem in quantum mechanics may look so deep and doomed, but one may be able to pull his bootstrap to get out of the desperation as Baron M\"unchhausen did. Following the ideas developed in large $N$ matrix models \cite{Lin:2020mme}\cite{Kazakov:2021lel}, they have pursued a new bootstrap approach to problems in quantum mechanics \cite{Han:2020bkb}\cite{Berenstein:2021dyf}\cite{Bhattacharya:2021btd}\cite{Aikawa:2021eai}\cite{Berenstein:2021loy}\cite{Tchoumakov:2021mnh}\cite{Aikawa:2021qbl}\cite{Du:2021hfw}\cite{Bai:2022yfv}. In this note, we will study a classical analogue of their quantum mechanical bootstrap.

Given a classical Hamiltonian $H(q,p)$, let us consider a microscanonical average of an observable $O(q,p)$:
\begin{align}
\langle O(q,p) \rangle = \frac{\int dp dq O(q,p)\delta(E-H)}{\int dp dq \delta(E-H)} \ ,
\end{align}
which is a function of the energy $E$.
Since the microcaonical ensemble is stationary under the Hamiltonian  time evolution, we immediately find
\begin{align}
\langle \{H,O(q,p) \} \rangle = 0 \ , \label{poisson}
\end{align}
where $\{H, O\}$ is the Poisson bracket.

From the definition of the microcanonical ensemble, we can also derive the following relation: 
\begin{align}
\langle H O(q,p) \rangle =  E \langle O(q,p) \rangle \ . \label{energy} 
\end{align}
The positivity of the measure leads to the inequality
\begin{align}
\langle O^* O \rangle > 0 \ . \label{positivity}
\end{align}
We may regard them as the classical analogue of the constraints they used in the bootstrap program in quantum mechanics \cite{Han:2020bkb}.  We here observe that taking $\hbar \to 0$ limit of their program reduces to the bootstrap study of the (classical) microcanonical ensemble of the same Hamiltonian.

As a simple example, consider $H =p^2 + x^2 + g x^4$. Studying $\langle \{H, x^s \} \rangle$, $\langle\{H, p x^s \} \rangle$, and $\langle H x^s \rangle$, we can derive the recursion relation
\begin{align}
4t E \langle x^{t-1} \rangle - 4(t+1) \langle x^{t+1} \rangle - 4g (t+2) \langle x^{t+3} \rangle = 0 \ . \label{recursion}
 \end{align}
The only difference from \cite{Han:2020bkb} is that the term $t(t-1)(t-2) \langle x^{t-3} \rangle $, which is proportional to $\hbar$, is missing. When $t=1$, there is no difference, so the relation $\langle x^4 \rangle = \frac{E-2\langle x^2 \rangle}{3g}$, which is essentially the Virial theorem, holds both in quantum mechanics and in the classical microcanonical ensemble.

\begin{figure}[tbh]
\centering
\includegraphics[width=7cm]{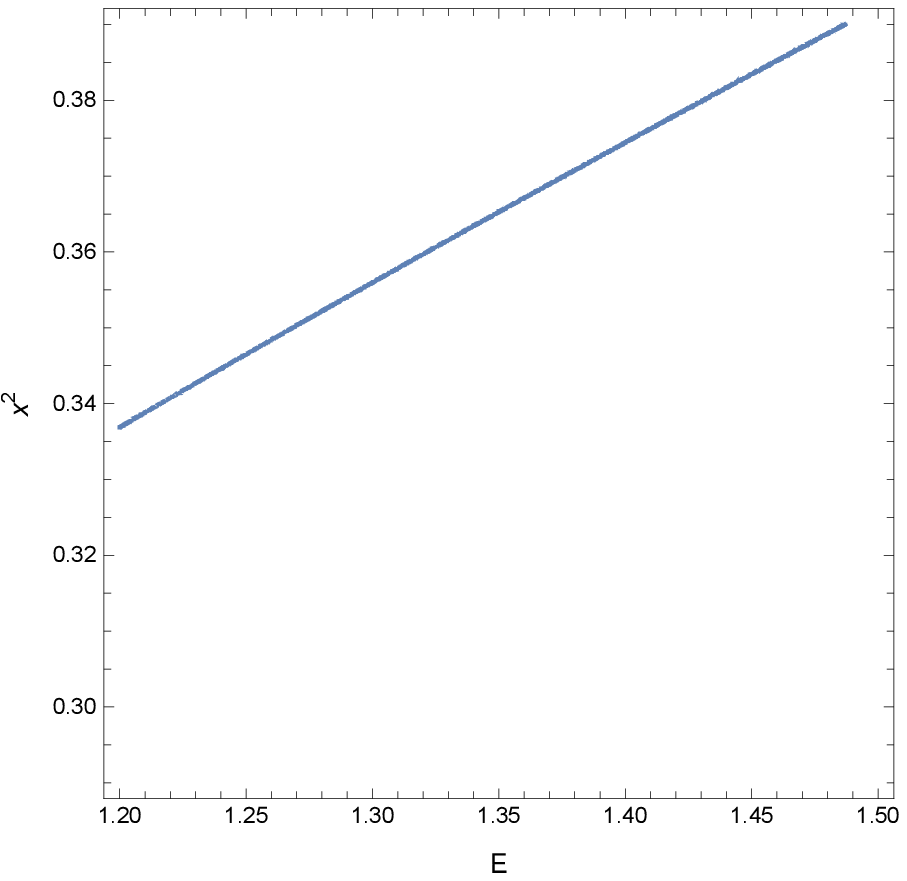}
\includegraphics[width=7cm]{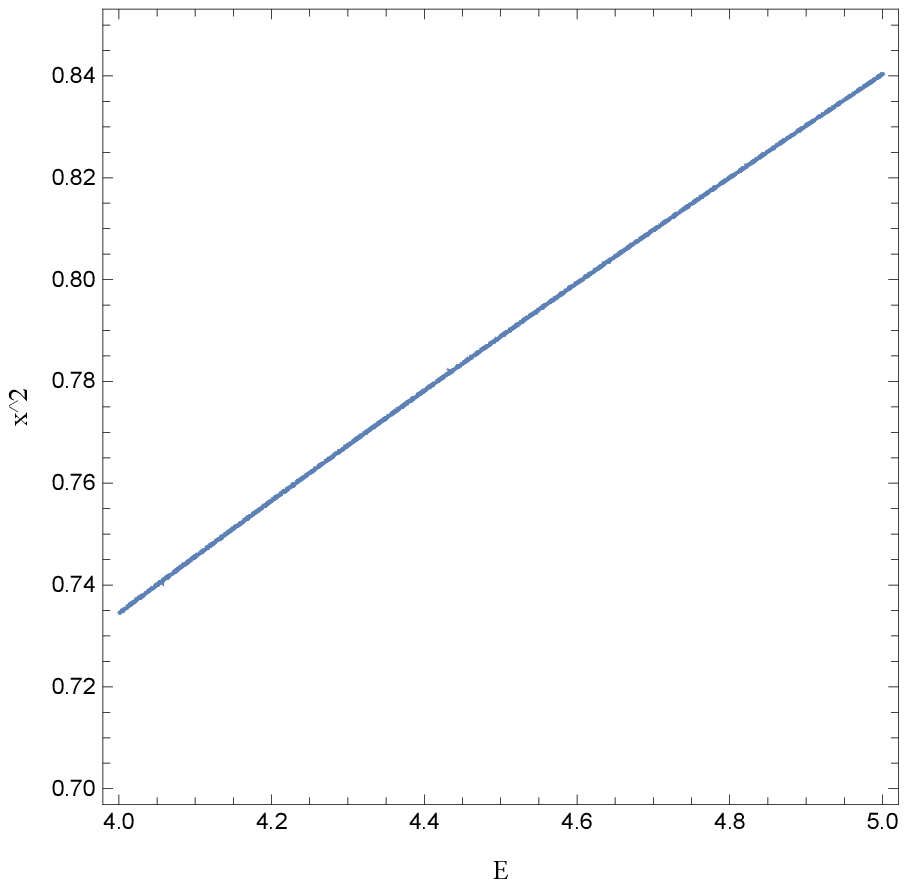}
\caption{Microcanonical bootstrap ($K=7$) for an anharmonic oscillator. Horizontal axis is $E$; vertical axis is $\langle x^2 \rangle$. The reason why we see tiny dots rather than a connected region is simply a lack of sampling points: the allowed region is extremely narrow.}
\label{fig1}
\end{figure}

\begin{figure}[tbh]
\centering
\includegraphics[width=7cm]{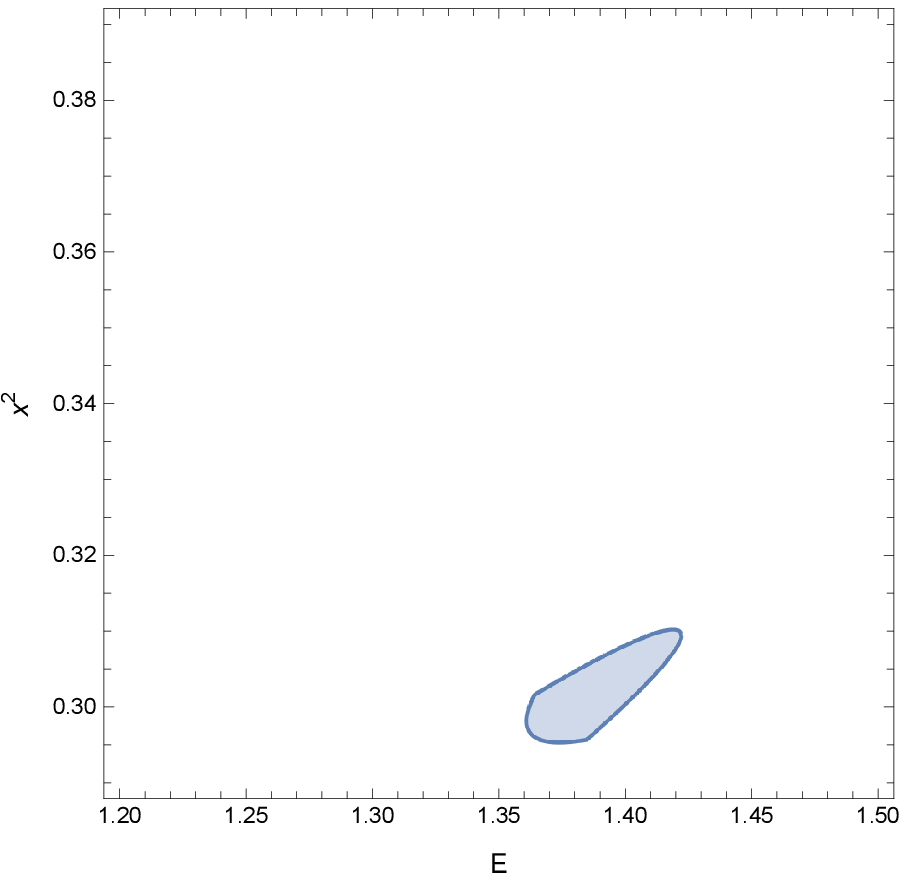}
\includegraphics[width=7cm]{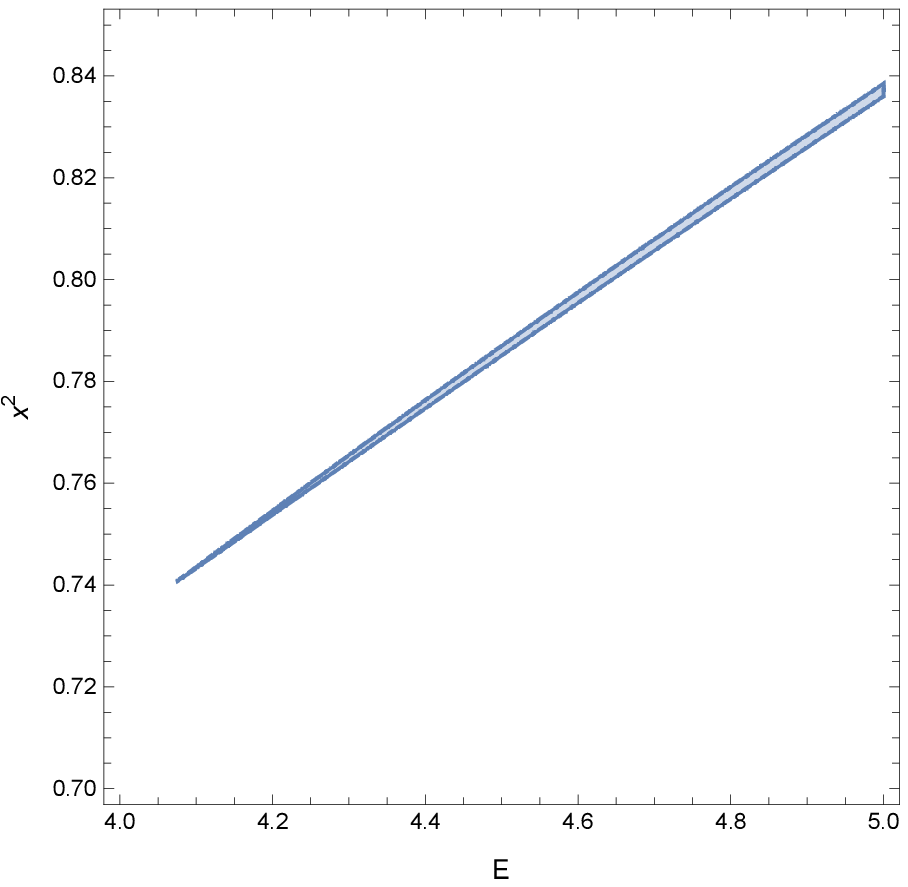}
\caption{Quantum mechanical bootstrap ($K=7$) for the anharmonic oscillator. We reproduced the results in \cite{Han:2020bkb}.}
\label{fig2}
\end{figure}

By using the recursion relation \eqref{recursion} we can express $\langle x^{n} \rangle$ $(n>2)$ in terms of $E$ and $\langle x^2 \rangle$ (as well as $\langle x \rangle =0$). Following \cite{Han:2020bkb}, we demand $\langle O^* O \rangle > 0$ with $O=\sum_i^K c_i x^i$ to constrain the bootstrap data (i.e. $E$ and $\langle x^2\rangle$). The results at $g=1$ and $K=7$ are shown in figure 1.\footnote{The numerical analysis was done by Mathematica. We could do the higher $K$, but finding the allowed region becomes more numerically difficult.}
Unlike in the quantum mechanical bootstrap, we find no archipelago, but we discover an extremely thin line that will limit to the  microcanonical ensemble curve 
\begin{align}
\langle x^2 \rangle(E) = \frac{\int dx \frac{x^2}{\sqrt{E-x^2-gx^4}}}{\int dx \frac{1}{\sqrt{E-x^2-gx^4}}}
\end{align}
with the exponential accuracy.\footnote{The integral could be evaluated by  complete elliptic integrals. Asymptotically, it is given by $\langle x^2 \rangle \sim \sqrt{E}$.}

For a comparison, we show the corresponding result in the quantum mechanical bootstrap with $K=7$, at which the islands first appear, in figure 2. With larger $E$, we see the entire archipelago of bootstrap (if we take large enough $K$). We also note that in the large $E$ limit, the recursion relation in the quantum mechanics becomes identical with that of the microcanonical ensemble as expected, and this is one of the reasons why the quantum mechanical bootstrap in the larger $E$ regime gives a narrow island  \cite{Han:2020bkb}.\footnote{In \cite{Lawrence:2021msm}, they proposed that minimizing energy over the allowed bootstrap data leads to more numerical efficiency by utilizing the semi-definite programming. In our case, minimizing the energy (at $E=0$) becomes trivial and exact already at $K=2$.}

Finally, we have one cautionary remark about what we mean by the ``microcanonical ensemble" when we have more than one dynamical degrees of freedom or when the phase space is disconnected. In the microcanonical ensemble, the distribution is completely specified by the microcanonical measure $dp^n dq^n \delta(E-H(q_i,p_i))$ and the expectation values of all the observables $\langle O(q_i,p_i) \rangle$ are fixed. However, our approach taken here (i.e. based on \eqref{poisson}, \eqref{energy} and \eqref{positivity}) does not specify the sub-distribution inside the fixed energy hypersurface, so we have a less predictive power than averaging over the true microcanonical measure. It may be related to the observation that the quantum mechanical bootstrap (in particular, the recursive technique) requires more creative thinking when we want to study more than one (effective) dynamical degrees of freedom.\footnote{A bootstrap approach to thermodynamic quantities will be found in \cite{toappear},}

As a simple demonstration of the above statement, let us consider the double-well potential: $H =p^2 - x^2 + g x^4$ at $g=1$. The recursion relation becomes
\begin{align}
4t E \langle x^{t-1} \rangle + 4(t+1) \langle x^{t+1} \rangle - 4g (t+2) \langle x^{t+3} \rangle = 0 \ , \label{recursion2}
 \end{align}
 and we demand $\langle O^* O \rangle > 0$ with $O=\sum_i^K c_i x^i$. The bootstrap result with $K=8$ is shown in figure 3. We here assume $\langle x \rangle =0$ for simplicity although in classical mechanics it is not necessary (in particular when $E<0$). The situation near (or below) $E=0$ is qualitatively different from what we have seen in the anharmonic oscillator. The region does not shrink even for larger $K$.

\begin{figure}[tbh]
\centering
\includegraphics[width=7cm]{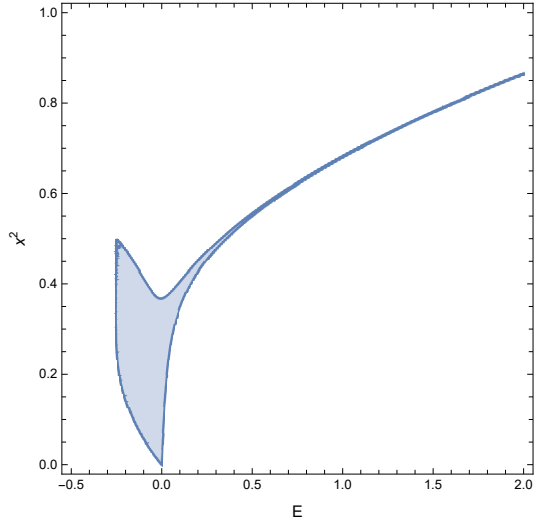}
\caption{Microcanonical bootstrap ($K=8$) for a double-well potential. It should be contrasted with the results in \cite{Bhattacharya:2021btd}\cite{Berenstein:2021loy}.}
\label{fig3}
\end{figure}

At $E=0$, we have three independent solutions $x=0$, $-2^{-1/2} <x<0 $, and $0<x<2^{-1/2}$. The microcanonical distribution would give a definite weight for each and give the unique prediction for $\langle x^2 \rangle$. Our bootstrap program, as it is\footnote{Our bootstrap analysis is chosen to be the classical limit of \cite{Bhattacharya:2021btd}\cite{Berenstein:2021loy}. When $E<0$, the lower boundary of the allowed region in figure 3 seems to be saturated by a ``tunneling solution" with an imaginary momentum. It is possible to shrink the region by demanding the positivity constraint coming from $p$. For example, $\langle p^2 \rangle >0 $ gives $\langle x^2 \rangle>-2E$, capping off   the region realized by the ``tunnelling solution".}, does not constrain a sub-distribution over them, necessarily giving rise to the larger allowed region for $\langle x^2 \rangle$. 

In order to see the effect of the sub-distribution, let us fix $E (<0)$ and $\langle x^2 \rangle$ and study the allowed region for $\langle x \rangle$ and $\langle x^3 \rangle$. As we see in figure 4, we find a linear region that can be interpreted as a different superposition of $-2^{-1/2} <x<0 $ and $0<x<2^{-1/2}$ solutions, the origin being the equal weight solution. This clearly shows that the classical limit of the quantum mechanical bootstrap may allow more solutions than the equal weight micro-canonical averaging.\footnote{In contrast, note that in quantum mechanics, $\langle x \rangle = \langle x^3 \rangle = 0$ for fixed $E$ and the analogue of figure 4 does not exist.}

In the globe, the classical limit of the archipelago could have been a peninsula in the ice age, and so  might be in the bootstrap.

\begin{figure}[tbh]
\centering
\includegraphics[width=7cm]{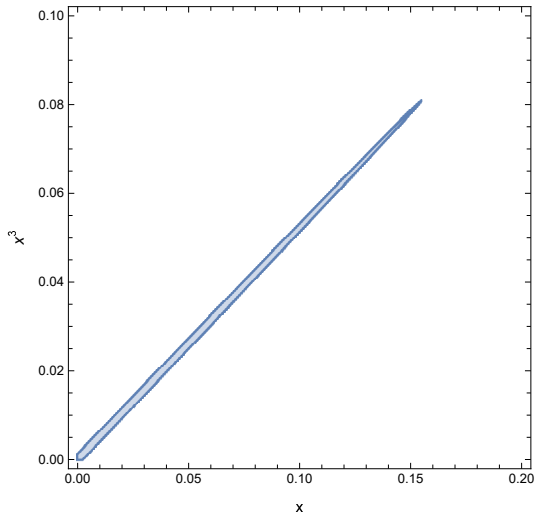}
\caption{Microcanonical bootstrap ($K=7$) for a double-well potential with fixed $E= -0.1$ and $\langle x^2 \rangle = 0.423$. The horizontal axis is $\langle x \rangle $ and the vertical axis is $\langle x^3 \rangle$.}
\label{fig4}
\end{figure}



\section*{Acknowledgements}
I would like to thank T.~Morita for the discussion at his poster session in YITP workshop Strings and Fields 2021, where I raised a concern about the thermodynamic predictions based on the microcanonical bootstrap. I would also like to thank Y.~Hatsuda for his lecture and the discussion. This work is in part supported by JSPS KAKENHI Grant Number 21K03581.

\end{document}